\documentclass[letterpaper, 10 pt, conference]{ieeeconf} 
\IEEEoverridecommandlockouts                              
\usepackage{lipsum}
\usepackage{amsmath}
\usepackage{amsthm}
\usepackage{algorithm}
\usepackage{url}

\usepackage{lipsum}% http://ctan.org/pkg/lipsum

\usepackage[noend]{algpseudocode}
\usepackage{algpseudocode}
\usepackage[bottom]{footmisc}
\usepackage{tikz}

\usetikzlibrary{shapes.arrows, fadings}
\usetikzlibrary{decorations.text}

\usepackage{cite}

\usepackage{mathtools}

\DeclarePairedDelimiter\floor{\lfloor}{\rfloor}
\usepackage[misc]{ifsym}
\newcommand{\corrAuthor}{$^{\textrm{\Letter}}$}
\newcommand{\MyComment}[1]{\hspace*{\fill}\Comment{\parbox[t]{.4\linewidth}{#1}}}

\usepackage{flushend}

\begin{document}

\title{\LARGE \bf
Optimized Memoryless Fair-Share HPC Resources Scheduling \\ using Transparent Checkpoint-Restart Preemption
}

\author{
    Kfir Zvi$^{1}$, Gal Oren$^{2}$\corrAuthor
    \thanks{
        $^{1}$K. Zvi is with the Department of Computer Science, Ben-Gurion University of the Negev, Be'er-Sheva, Israel
        \& Israel Atomic Energy Commission, P.O.B. 7061, Tel Aviv, Israel
        \newline
        {\tt\small  zvikf@post.bgu.ac.il}
    }
    \thanks{
        $^{2}$G. Oren is with the Department of Computer Science, Technion – Israel Institute of Technology, Haifa 32000, Israel
        \& the Scientific Computing Center, Nuclear Research Center-Negev, P.O.B. 9001, Be'er-Sheva, Israel
        \newline
        {\tt\small  galoren@technion.ac.il}
    }

    \thanks{
        \newline
        * This work was supported by Pazy grant 226/20. Computational support was provided by the NegevHPC project~\cite{negevhpc}. The authors would like to thank Matan Rusanovsky, Harel Levin, Hagit Attiya and Danny Hendler for their fruitful comments.
        \newline
    }
}

\maketitle

\begin{abstract}
Common resource management methods in supercomputing systems usually include hard divisions, capping, and quota allotment. Those methods, despite their ‘advantages’, have some known serious disadvantages including unoptimized utilization of an expensive facility, and occasionally there is still a need to dynamically reschedule and reallocate the resources. Consequently, those methods involve bad supply-and-demand management rather than a free market playground that will eventually increase system utilization and productivity. In this work, we propose the newly Optimized Memoryless Fair-Share HPC Resources Scheduling using Transparent Checkpoint-Restart Preemption, in which the social welfare increases using a free-of-cost interchangeable proprietary possession scheme. Accordingly, we permanently keep the status-quo in regard to the fairness of the resources distribution while maximizing the ability of all users to achieve more CPUs and CPU hours for longer period without any non-straightforward costs, penalties or additional human intervention.    
\end{abstract}

\section{Introduction}
Demand for High-Performance Computing (HPC) resources has been skyrocketing over the last decade~\cite{meuer2014top500}. Consequently, said resources became more available over the cloud, affordable (due to competition~\cite{raas}), and immediately supplied~\cite{buyya2018manifesto,netto2018hpc,sadooghi2015understanding}.
Nevertheless, HPC on the cloud is limited~\cite{ferretti2019cloud, breuer2019petaflop}: (1) Cloud hardware is general purposed, and not tailor-made like a supercomputer. Hence, it is insufficient for large-scale coupled computations; (2) The current commercial tariff paradigm makes it more expensive to run continuous computations on the cloud, compared with on-premise resources. Thus, many R\&D institutions still establish in-house supercomputing facilities~\cite{top500}. However, unlike cloud resources, which are dynamically assigned according to an economic model~\cite{cui2020survey}, supercomputing facilities are usually assigned by hard divisions or by utilization capping.
These models reflect entitlements of \textit{entities} (groups or individual users)~\cite{kyaw2020scheduling},
that originate in centrally-managed work-plans. An entity expects immediate access to its resources, 
similarly to reserved cloud instances,
but lacking an economic mechanism -- which would have regulated user requests -- congestion is formed. To prevent group starvation, the congestion is usually mitigated by capping usage, static allocation and rationing CPU hours. 
Jobs are prioritized in queues, and low priority jobs may starve, in full accordance with the users' expectations.
These methods suffer from under-utilization of the expensive, divided facility, and from a frequent need 
to manually adjust the entitlements~\cite{uruchi2020game}. Therefore, there is a need for a scheduling mechanism that will alleviate  the barriers that prevent full resource pooling, and yet maintain the entities' expectation of \textit{fairness}:
Each entity gets at least its entitlement of CPUs when it has the suitable workload for it~\cite{dolev_no_2011}.

\section{The Optimized Memoryless Fair-Share}

The scheduling and resource allocation is usually performed using schedulers such as SLURM~\cite{slurm,yoo2003slurm,top500}.

A recent development of the SLURM scheduler is the seamless incorporation of the Distributed MultiThreaded Checkpointing (DMTCP)~\cite{ansel2009dmtcp} library, enabling it to transparently Checkpoint and Restart (C/R) a single-host, 
parallel or distributed computation. It does so in user-space, with no modifications to user code or the 
operating system, supporting a variety of HPC languages and infrastructures, including MPI and OpenMP~\cite{rodriguez2019job}.

Following this recent capability, we propose the Optimized Memoryless Fair-Share HPC Resource Scheduling (OMFS) method by enabling transparent resource pooling. 
Each entity is still guaranteed its \textit{CPU entitlement}, while not being limited to certain compute nodes via static allocation, and without computation capping. Instead, underutilized resources are assigned to entities with active demand. \textit{Non-preemptible} jobs can only be run up to the entity's entitlement. To utilize more resources,
jobs must be either \textit{preemptable} (may be killed) or \textit{checkpointable} (can C/R using DMTCP or otherwise). Hence, if an entity does not utilize its full entitlement, it can always increase its utilization up to its entitlement by evicting jobs of entities utilizing more than their allotment. If these jobs are checkpointable, they are transparently checkpointed prior to the eviction, and later restarted. 

\begin{algorithm*}[ht]
\begin{algorithmic}[1]
\normalsize
\Procedure{System Init}{$Users$, $Jobs_{Submitted}$, $Jobs_{Running}$}
\MyComment{Performed once at startup}
\label{sys_init_start}
\State $CPU_{Total} \gets N$ \label{cpu_total_def} \MyComment{Amount of CPUs in the entire system}

\State $CPU_{Idle} \gets 0$ \MyComment{Amount of idle CPUs in the entire system}

\State $Users \gets Users$ \MyComment{Predefined group of system users}

\State $Jobs_{Submitted} \gets Jobs_{Submitted}$ \label{jobs_queues_start} \MyComment{A predefined priority queue}
\State $Jobs_{Running} \gets Jobs_{Running}$ \label{jobs_queues_end} \MyComment{A predefined priority queue}
\State \textbf{\textit{foreach}} user $u \in Users$:
\State \hspace{0.5cm} $u.percent \gets p : p \in [0,100]$ \MyComment{Percentage of CPU allocation per user}
\State \textbf{assert} $\sum_{u \in Users}(u.percent) \leq 100$ \MyComment{Upper bound of allocations' percentages} \label{sys_init_end}
\EndProcedure

\item[]
\Procedure{Job Init}{User u, Job j} \MyComment{Performed for every job creation} \label{job_init_start}

\State $j.priority \gets pr : pr \in N_{0}$ \MyComment{Priority among the jobs of the user only}
\State $j.CPU_{Count} \gets cpus : cpus \in \{x \in N_{0}:0\le x\le N\}$ \MyComment{Amount of requested job's CPUs}
\State $j.user \gets u$ \label{job_init_end}
\EndProcedure

\item[]
\Procedure{Memoryless Fair-Share Scheduler()}{} \label{scheduler_start}
\State \textbf{\textit{while}} True: \MyComment{Continuously try to run jobs from $Jobs_{Submitted}$}
\State \hspace{0.5cm} $J \gets Jobs_{Submitted}.dequeue()$ 
\State \hspace{0.5cm} \textsc{Memoryless Fair-Share Runner}(J) \label{scheduler_end}

\EndProcedure

\item[]
\Procedure{Memoryless Fair-Share Runner}{\textit{Job J}} \label{runner_start}

\State $User_{PAbleJobsCPUCount} \gets \sum_{j \in \{ j \in Jobs_{Running} : j.user = J.user \textit{ \textbf{and} j \textbf{is} preemptable}\}} (j.CPU_{Count})$ \newline \MyComment{Amount of CPUs occupied by user's preemptable jobs}
\State $User_{NonPAbleJobsCPUCount} \gets \sum_{j \in \{ j \in Jobs_{Running} : j.user = J.user \text{\textit{ \textbf{and} j \textbf{is not} preemptable}\}}} (j.CPU_{Count})$ \newline \MyComment{Amount of CPUs occupied by user's non-preemptable jobs}
\State $User_{TotalJobsCPUCount} \gets User_{PAbleJobsCPUCount} + User_{NonPAbleJobsCPUCount}$
\State $User_{EntitledCPUCount} \gets \floor{(J.user.percent / 100)\cdot CPU_{Total}}$

\State \textbf{\textit{if}} $J$ \textit{\textbf{is not} preemptable \textbf{and}} $User_{NonPAbleJobsCPUCount} + J.CPU_{Count} \ge User_{EntitledCPUCount}$: \newline \MyComment{If the non-preemptable jobs exceed the user's entitlement} \newline
\State \hspace{0.5cm} $Jobs_{Submitted}.enqueue(J)$ \MyComment{Do not allow to run}
\State \hspace{0.5cm} \textbf{\textit{return}}
\State \textbf{\textit{elif}} $CPU_{Idle} > J.CPU_{Count}$: \MyComment{If enough resources are available}
\State \hspace{0.5cm} \textbf{\textit{goto}} \ref{schedule_job} \MyComment{Allow to run anyways}
\State \textbf{\textit{elif}} $J.CPU_{Count} > User_{EntitledCPUCount} - User_{TotalJobsCPUCount}$: \newline \MyComment{Check job's CPU request fits within entitlement}
\State \hspace{0.5cm} $Jobs_{Submitted}.enqueue(J)$ \MyComment{No fit, the job remains in $Jobs_{Submitted}$}
\State \hspace{0.5cm} \textbf{\textit{return}}
\State \textbf{\textit{else}}: \MyComment{User is entitled to run \textit{J}, so make resources available} \newline
\State \hspace{0.5cm} \textbf{\textit{while}} $CPU_{Idle} < J.CPU_{Count}$: \MyComment{Until reaching the entitlements}
\State \hspace{1.0cm} $Job_{checkpoint} \gets Jobs_{Running}.dequeue()$ \MyComment{Checkpoint the least prioritized running jobs}
\State \hspace{1.0cm} \textbf{\textit{if}} $Job_{checkpoint}$ \textit{\textbf{is} checkpointable}: \MyComment{If it is not checkpointable, drop it}
\State \hspace{1.5cm} $Jobs_{Submitted}.enqueue(Job_{checkpoint})$ \MyComment{Checkpointed job back to $Jobs_{Submitted}$}
\State \hspace{1.0cm} $CPU_{Idle} \gets CPU_{Idle} + Job_{checkpoint}.CPU_{Count}$ \MyComment{Update amount of idle CPUs in the system}
\State $Jobs_{Running}.enqueue(J)$ \label{schedule_job} \MyComment{\textit{J} is eligible to run. Schedule it}
\State $CPU_{Idle} \gets CPU_{Idle} - J.CPU_{Count}$ \MyComment{Update amount of idle CPUs in the system} \label{runner_end}
\EndProcedure

\caption{Optimized Memoryless Fair-Share HPC Resource Scheduling using Transparent Checkpoint-Restart Preemption}
\label{algo}
\end{algorithmic}
\end{algorithm*}

In Algorithm \ref{algo} we describe a simplified algorithm, which is based on pre-existing priority queues. They can be governed by any prioritization policy such as FIFO or priority-by-user (lines \ref{jobs_queues_start}--\ref{jobs_queues_end}).
The algorithm starts with system initialization, which occurs only once at startup (lines \ref{sys_init_start}--\ref{sys_init_end}), and with a job initialization procedure which is run every time a new job is initialized (lines \ref{job_init_start}--\ref{job_init_end}).
The \textsc{Memoryless Fair-Share Scheduler} in lines \ref{scheduler_start}--\ref{scheduler_end} iterates over the queue, trying to execute submitted jobs using the main algorithm of \textsc{Memoryless Fair-Share Runner} (lines \ref{runner_start}--\ref{runner_end}). The \textsc{Memoryless Fair-Share Runner} may run the job and may also preempt other jobs, according to the fairness policy described above.

This preemption method lets underutilized entitlements be temporarily used by other entities, and thus improves the utilization over a capping-based system.
Furthermore, an entity can use it to run a single job that is larger than its whole entitlement, without any manual intervention.

Recurrent C/R preemption operations may lead to thrashing.
To resolve this problem we take two approaches: 
First, each job is allowed a \textit{quantum}: A minimal running duration (e.g., 30 minutes). 
The running jobs queue demotes jobs that have been running uninterruptedly for at least a quantum. 
Enlarging the quantum lowers the frequency of C/R operations at the price of non-instantaneous execution of accredited jobs. 
Second, we utilize current developments in distributed file systems~\cite{anderson2019assise, yang2019orion} over non-volatile memory~\cite{izraelevitz2019basic} to reduce the cost of C/R operations: These memory modules are connected to the CPU via the Memory Bus, and can be accessed faster than standard storage.
We intend to use Persistent Memory File Systems~\cite{liu2020survey} (e.g., SplitFS~\cite{splitfs2019}, NOVA-Fortis~\cite{nova2017}, or Assise~\cite{assise2020}) over the Intel Optane™ DataCenter Persistent Memory 
(DCPMM) module~\cite{izraelevitz2019basic}, so that DMTCP will be able to use it transparently. A better approach to resolve this problem is to perform source code modifications to DMTCP, to support Direct Access (DAX) to persistent memory via libraries such as PMDK~\cite{scargall2020introducing}. DAX allows accessing the persistent memory directly from the application. This way, we can utilize the DCPMM to the maximum extent to reduce thrashing cost, as it will serve as a fast non-volatile storage (essential property for checkpointing) and as a memory from which the restart operation could directly start from.  

\section{Related Work}

\subsection{Technical scheduler and transparent C/R}
We follow Rodriguez et al.~\cite{rodriguez2019job} in basing our work on 
SLURM because it is the only scheduler which supports C/R in the form of an API. 
DMTCP is a system-level C/R library which allows to perform C/R operations without any source code modifications; it supports SysV enhancements such as System V shared memory~\cite{dmtcp_faq_2021} which many MPI implementations employ; and it supports InfiniBand~\cite{cao_transparent_2013}. This is in contrast to BLCR~\cite{hargrove2006berkeley}, which requires re-compilation and possibly re-tuning of the kernel module and does not support neither SysV enhancements nor InfiniBand, and CRIU~\cite{criu}, which currently does not even have support for parallel or distributed applications. Furthermore, Rodriguez et al.~\cite{rodriguez2019job} provided two sample scheduling algorithms based on their work. These algorithms do not share the same goal as ours -- which is mainly to increase system utilization -- as one is used to reduce power consumption by scheduling jobs on as least nodes as possible, and the other is used to schedule jobs with higher priority on better hardware.

\subsection{Scheduling methods in SLURM}
Currently, SLURM supports two main scheduling methods: \textit{sched/builtin} which is a simple FCFS algorithm, and \textit{sched/backfill} which is an extension to the basic FCFS algorithm that enables scheduling of jobs that are not at the head of the queue, as long as they do not cause a delay in the scheduling of reserved jobs (jobs which their starting time has already been determined).
Both FCFS and backfill have been studied thoroughly and while being relatively simple, they are used extensively in current supercomputers' job scheduling systems. The main reason is that practical limitations prevent the use of other scheduling algorithms~\cite{Niu2012EmployingCT}. In particular, FCFS-backfill relies on users' estimates for jobs' run-time, which have been proven to be highly inaccurate~\cite{Niu2012EmployingCT, mualem1998, lee2004user}. This evidently leads to very poor efficiency for scheduling systems~\cite{Niu2012EmployingCT}. Niu et al.~\cite{Niu2012EmployingCT} have also shown an example of how checkpointing along with preemptive scheduling can increase the performance of the backfill algorithm, and in this work we follow this approach in practice. To the authors' knowledge, aside from \cite{rodriguez2019job}, this has never been done before.

\subsection{Fairness}
Among others, \textit{fairness} is a common attribute of a job scheduling system in HPC, and it is of utmost importance for HPC users, maybe even more than the productivity of the HPC system~\cite{yuan_pv-easy_2010}.
The concept of \textit{fairness} is well established~\cite{henry_unix_1984}. As \textit{fairness} and performance are in conflict in parallel and distributed job scheduling, previous works either focused on providing fair scheduling or improving performance~\cite{yuan_pv-easy_2010}. A stronger notion of \textit{fairness}, which we did not implement, is \textit{strict fairness}, where no job is delayed by any other job of lower priority.

Our scheduler implements \textit{memoryless fairness}, as opposed to \textit{history-based fair-share}. Schedulers which use this concept re-prioritize jobs according to previous job metrics, with the purpose of reaching a designated level of these metrics (e.g., mean usage or wait time) over a time frame, such that it reflects its entitlements. \textit{History-based fair-share} is less predictable and transparent than \textit{memoryless fair-share}.
As implemented in popular schedulers~\cite{hoopes_slurm_2019, moab_fairshare2019},
\textit{history-based fair-share} is only sensitive to jobs that have been completed and logged by the rolling fair-share calculation.
If a group submits a bunch of jobs quickly, they could potentially fill the entire system with their jobs~\cite{steffen2019better}. Furthermore, since SLURM's implementation of fair-share~\cite{hoopes_slurm_2019, slurm_classic_fairshare} uses a decay factor\footnote{SLURM's fair-share design includes an accounts hierarchy, in which each account is given a normalized relative share of its parent account. For each account, SLURM keeps track of its \textit{actual usage}, and adjusts its \textit{effective usage} accordingly so it represents the account's shares according to the fair-share policy. Both \textit{actual usage}, and \textit{effective usage} are affected by a predefined decay factor, which gives higher weight to a more recent usage of the resources made by each account.}, it is hard to predict when exactly a submitted job will be scheduled.

\section{Conclusion and future work}
In this work we presented the \textit{Optimized Memoryless Fair-Share HPC Resources Scheduling using Transparent Checkpoint-Restart Preemption} (OMFS).
We permanently keep the status quo regarding the \textit{fairness} of the resources distribution -- which is crucial to the users’ work-fashion -- while maximizing the ability of all users to achieve more CPUs and CPU hours for longer periods of time without any non-straightforward costs, penalties or additional human intervention.

In accordance with the notions presented in this paper, we intend to test our scheduler on the NegevHPC~\cite{negevhpc} semi-supercomputer cluster. The first step will be to use SLURM's API using PySlurm~\cite{pyslurm} to build our scheduler, and to evaluate users' behaviour. Next, we plan to build a dedicated SLURM plugin, based on Rodriguez et al.~\cite{rodriguez2019job} work, and to apply our solution to the thrashing problem using non-volatile memory.

\bibliographystyle{plain}
\bibliography{bibliography}

\end{document}